\documentclass[sigconf, nonacm]{acmart}

\usepackage{caption}
\usepackage{subcaption}
\usepackage{multirow}
\usepackage{makecell}

\newcommand{\agr}{\textcolor{black}{\emph{Agreement}}}
\newcommand{\sha}{\textcolor{black}{\emph{Intent to Share}}}

\newcommand{\cond}{\textcolor{black}{\textsc{XAI Condition}}}
\newcommand{\vera}{\textcolor{black}{\textsc{News Veracity}}}

\AtBeginDocument{
  \providecommand\BibTeX{{
    \normalfont B\kern-0.5em{\scshape i\kern-0.25em b}\kern-0.8em\TeX}}}

\begin{document}

\title{XAI in Automated Fact-Checking? The Benefits Are Modest and There's No One-Explanation-Fits-All}

\author{Gionnieve Lim}
\email{gionnievelim@gmail.com}
\orcid{0000-0002-8399-1633}
\affiliation{
  \institution{Singapore University of Technology and Design}
  \streetaddress{8 Somapah Rd}
  \country{Singapore}
  \postcode{487372}
}

\author{Simon T. Perrault}
\email{perrault.simon@gmail.com}
\orcid{0000-0002-3105-9350}
\affiliation{
  \institution{Singapore University of Technology and Design}
  \streetaddress{8 Somapah Rd}
  \country{Singapore}
  \postcode{487372}
}

\renewcommand{\shortauthors}{Lim and Perrault}

\begin{abstract}
The massive volume of online information along with the issue of misinformation has spurred active research in the automation of fact-checking. Like fact-checking by human experts, it is not enough for an automated fact-checker to just be accurate, but also be able to inform and convince the user of the validity of its predictions. This becomes viable with explainable artificial intelligence (XAI). In this work, we conduct a study of XAI fact-checkers involving 180 participants to determine how users' actions towards news and their attitudes towards explanations are affected by the XAI. Our results suggest that XAI has limited effects on users' agreement with the veracity prediction of the automated fact-checker and on their intent to share news. However, XAI nudges users towards forming uniform judgments of news veracity, thereby signaling their reliance on the explanations. We also found polarizing preferences towards XAI and raise several design considerations on them.
\end{abstract}

\ccsdesc[500]{Human-centered computing~Empirical studies in HCI}

\keywords{misinformation, automated fact-checking, explainable artificial intelligence, interpretable machine learning, human-centered design, human-AI interaction}

\maketitle

\section{Introduction}
Misinformation is one of the key challenges of social media. To address it, social media firms have partnered with professional fact-checkers to assess content on their platforms.
Through fact-checking by experts, platform content is reviewed and rated for their accuracy and may eventually be removed if found in breach of company standards~\cite{Bettadapur2020, Facebook2021, Mantas2021}. While effective as a measure against false content, the sheer volume of information that is created every moment is beyond the capabilities of the limited number of expert fact-checkers, making the scalability of fact-checking a fundamental concern~\cite{Moy2021}. In seeking to address this, researchers have explored automated fact-checking.
Automated fact-checking uses machine learning algorithms and artificial intelligence to assess online content and research has focused on developing various models to improve the detection accuracy of false content~\cite{Islam2021, Khan2021, Sharma2019, Wu2019}. There has also been research seeking to make the models more understandable to humans through the work of interpretable machine learning (IML) and explainable artificial intelligence (XAI)~\cite{Zhou2020}.

\begin{figure*}[!htb]
  \centering
  \includegraphics[width=\linewidth]{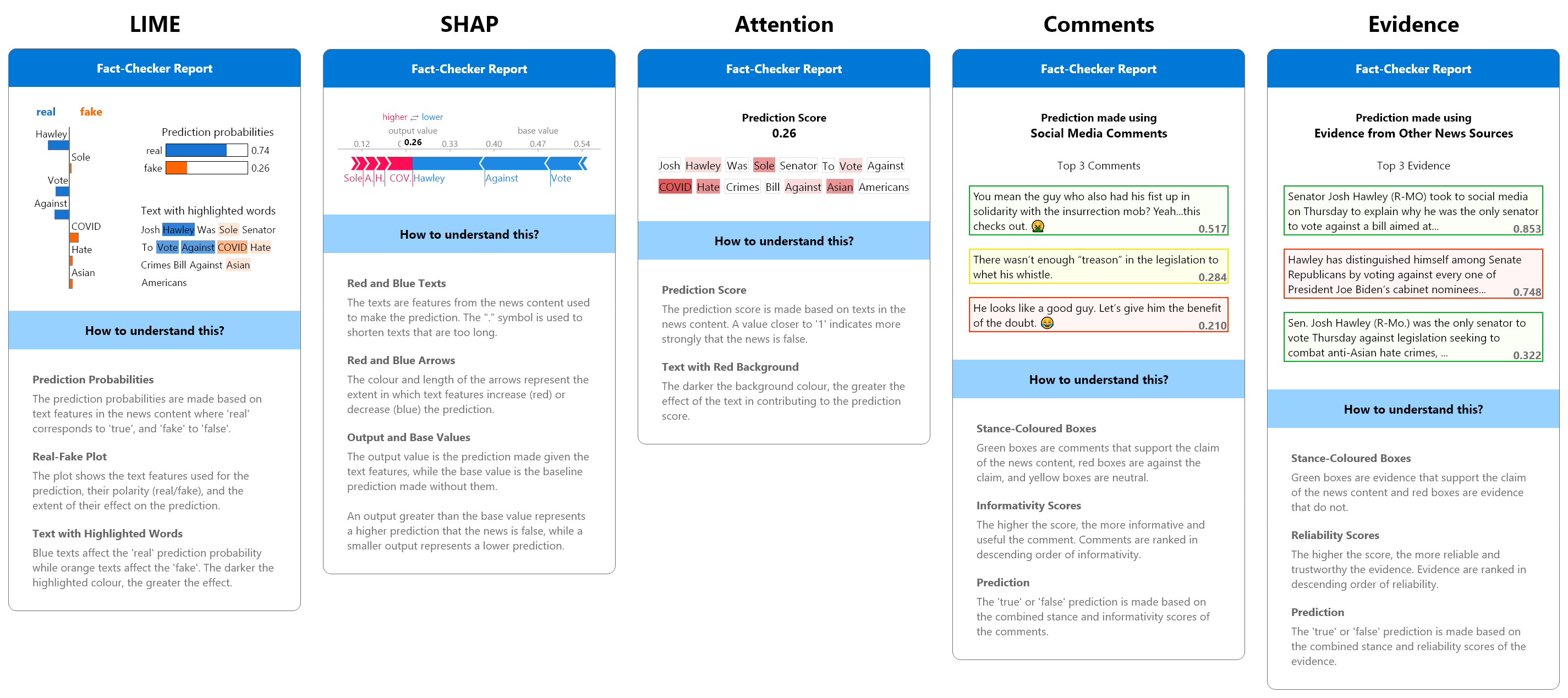}
  \caption{The XAI conditions for the automated fact-checker. The fact-checker reports, each with a "How to understand this?" guide, are based on the TRUE news shown in Figure~\ref{fig:NewsPrediction}. The LIME, SHAP and Attention conditions display similar data with a graphical visualization while the Evidence and Comments conditions use a textual visualization.}
  \label{fig:XAI}
\end{figure*}

With ethical advances in the technology discourse urging for greater human-centered design in commercial machine learning and artificial intelligence applications, social media giants are shifting towards Responsible AI in misinformation detection. With this, it has become increasingly important to assess the effectiveness of automated fact-checkers and the practicality of explanations. In this paper, we ran an experimental study with 180 participants comparing five popular XAI interfaces sourced from the literature (LIME~\cite{Ali2021}, SHAP~\cite{Reis2019}, Attention~\cite{Hansen2019}, Comments~\cite{Tian2020}, Evidence~\cite{Popat2018}) (Figure~\ref{fig:XAI}). We sought to understand how XAI can benefit automated fact-checking in terms of the user's agreement with the fact-checker on the veracity of the news and on their intent to share them. The perceptions that users have towards the XAI interfaces were also investigated. Our results suggested that while XAI interfaces have mixed effects on agreement and limited effects on the intent to share, they are relied upon in determining the veracity of news. We also found that users had polarizing preferences towards different types of XAI interfaces based on the types of visualizations and information shown. We identified five qualities that were commonly considered when participants assessed the explanations and raise several usability considerations on them.

The contribution of this work is two-fold:
\begin{itemize}
    \item We advance efforts in social computing on the automation of the fact-checking process, particularly identifying that XAI has modest benefits by increasing the reliance on the veracity predictions provided by the automated fact-checker but has limited effects on users' overall veracity judgment and intent to share news.
    \item We also advance the literature on human-AI interaction by presenting a set of design considerations and a prospective user evaluation framework on explanations which, while explored in the context of fact-checking, would be relevant in other contexts given that the explanations investigated are common across other domains.
\end{itemize}

\section{Related Work}

Our work is situated in the domains of misinformation and human-centered AI. We discuss the intersection of these domains and how XAI techniques have been proposed for fact-checking, highlighting those of interest in our study.

\subsection{Misinformation on Social Media}
Researchers have tracked the spread of misinformation on social media following major events like elections~\cite{Allcott2019}, disasters~\cite{Wang2018b} and pandemics~\cite{Mejova2020}. The greatest sources of misinformation may be attributed to just a few groups and people~\cite{KF2018}. Superspreaders such as influencers, public figures, and hyper-partisan media outlets that have a large follower base post false content repeatedly and across several platforms, enabling misleading narratives to spread quickly among and across their unassuming audiences and to their cascading networks~\cite{CIP2021, Bovet2019, Jasser2019}. Social bots which are automated accounts that interact on social media in a human-like manner also amplify misinformation through the sharing of false content, by replying to posts, and using mentions to induce greater engagement~\cite{Himelein2021, Julian2020, Wang2018}. The rise of large language models capable of rapidly producing realistic content has also led to concerns over its potential misuse to generate false content~\cite{DeAngelis2023}.

With these contributing factors, misinformation on social media has become a key challenge of the 21st century. In response to this challenge, automated solutions have been deployed on social media platforms to detect false content and to directly remove the content or flag them for verification by professional fact-checkers~\cite{Bettadapur2020, Facebook2021, Mantas2021}. Automated fact-checking have also been used to verify and highlight false content~\cite{Factinsect}. Yet, the complexity of these models make them difficult to understand, raising ethical concerns on their application and consequences.

\subsection{Human-Centered AI}
With leaping advances in AI and the application of these technologies in numerous domains, ethical concerns have been raised about the fairness, transparency and accountability of the opaque boxes~\cite{Shneiderman2021}. Human-centered AI is a sociotechnical field seeking to amplify human agency by enhancing the human understanding of the internal workings of AI systems and empowering people to make better-informed decisions. In contrast to data-centric AI that draws on statistical patterns in big data and machine learning to make decisions in ways that are incomprehensible and inaccessible to humans, human-centered AI seeks to build AI capable of explaining its decisions and is reconcilable with the moral and ethical standards of society~\cite{AI4EU2020}.

Explainable AI is one such initiative~\cite{Gunning2019}. By building a transparent model over an underlying opaque model, explainable AI aims to describe the internal mechanics of the model by explaining the features that attribute to the model's prediction. Another closely related concept often used interchangeably with XAI is IML that considers the extent to which a cause and effect can be observed directly in the underlying model. There has been active research into human-centered and explainable AI.
One aspect examines how people understand, interact with, accept, and trust XAI~\cite{Anik2021, Yin2019, Weitz2019, Ghai2021}, where we place our work. XAI has been used in many domains with several benefits to Human-AI interaction and we seek to extend this investigation in the domain of misinformation that stands to benefit from explanations to assist users in understanding the automated fact-checkers' decision-making processes when determining the veracity of content.

\subsection{Explainable Automated Fact-checking}
Social media platforms like Facebook, Twitter and TikTok have been working towards a healthier information ecosystem by incorporating deep learning technologies to their products to moderate content through flagging, labeling and removal~\cite{Twitter2020, Facebook2020, Tiktok2020}. In mid 2021, the companies announced their intentions towards Responsible ML and AI, in terms of the fairness and transparency of their algorithms~\cite{Facebook2021b, Twitter2021}. Among their plans is to build XAI solutions that inform users on how the platform algorithms work, and the impact of the algorithms on the platform experience.

In the context of automated fact-checking that aims to predict whether a piece of content is real or fake, there is already growing research on XAI fact-checkers for media like texts, images and videos. Two popular techniques for explainability are the local interpretable model-agnostic explanations (LIME)~\cite{Ribeiro2016} and Shapley additive explanations (SHAP)~\cite{Lundberg2017}.
There has also been considerable interest in the attention mechanism that places weights on representations that can be visualized to provide interpretability to text-based models~\cite{Cruz2020, Aloshban2020, Borges2019, Hansen2019, Redi2019}. The attention mechanism has also been adapted in various ways to identify top social media comments that help to explain the prediction~\cite{Shu2019, Tian2020}, and in a similar vein, top reliable evidence from other sources~\cite{Popat2018, Nguyen2018}.

While there are works comparing different types of explanations~\cite{vanderWaa2021, Wang2021}, it is unclear how explanations are understood and may influence users' behavior in the automated fact-checking context. Focusing on common techniques in explainable misinformation detection, we investigate how predicted veracity labels accompanied by the explanations that are applied to problematic posts on social media affect people's perceived veracity of news and their intent to share them.

\section{Method}
To understand the effects of different automated fact-checking explanations, we conducted a between-subjects experiment with 180 participants. Informed by prior literature~\cite{Clayton2020, Kirchner2020, Pennycook2019}, we assessed the effects by considering the participants' agreement with the fact-checker on the veracity of the news and their intent to share them.

The study sought to answer the following research questions (RQs):
\begin{enumerate}
    \item What are the effects of different explanations on users' agreement with the fact-checker on the veracity of news?
    \item What are the effects of different explanations on users' intent to share news?
    \item What are the preferences towards explanations in automated fact-checking?
\end{enumerate}

\subsection{Analysis and Comparison}
\label{sec:faircomparison}
We investigated the most common types of XAI techniques in misinformation detection literature by systematically scanning the literature. Our search surfaced two main categories: feature-based techniques where explanations have graphical representations of weighted features (the LIME, SHAP and Attention conditions) and example-based techniques where explanations draw on textual data from other sources (the Comments and Evidence conditions) (Figure~\ref{fig:XAI}). To compare our results, we also included a \emph{Control} condition where no explanation was provided.

We assessed the main effects of the XAI interfaces by analyzing all five conditions and the control. For a finer discussion, we focused on the comparison between techniques of the same graphical or textual categories as comparing across categories would have been less meaningful. For instance, we can be confident that the reason was due to the difference in the visualization for a difference between LIME and SHAP (both graphical), whereas if the difference was between LIME (graphical) and Evidence (textual), it would be difficult to tell whether it was due to the visualization or the information presented.

\subsection{Procedure}
We collected the data using an online web app created by the researchers.
Participants were recruited on Amazon Mechanical Turk (MTurk) through Human Intelligence Tasks (HITs). Participation was voluntary and it was possible to withdraw at any time with no data being collected. The study was approved by the University's Institutional Review Board (IRB) and participants were reimbursed for their time at a rate of 7.25 USD/hour as per our IRB guidelines.

When participants accepted the HIT, they were given a link to the web app to complete a screener survey. They would be redirected to the study upon passing. There were 12 links for the random redirection, six for each XAI condition and the control, and another six with counterbalanced Likert scales. Participants who did not pass were advised to return the HIT instead.

After filling their demographic details, participants proceeded to the experiment. A walkthrough containing descriptions of the news, automated fact-checker, and the task instructions were provided. Participants then proceeded to the task of rating a randomized set of 12 news with the assigned XAI interface. For each news, they provided their perceived veracity of the news (4-points, False to True) and their intent to share it (5-points, Definitely Not to Definitely Yes). Thereafter, they completed a post-experiment survey containing an open-ended question on their perceptions towards the assigned XAI interface.
Upon completion, participants were given instructions to generate a unique completion code to submit on MTurk.

\subsection{Materials}

\subsubsection{Stimuli/News}
\begin{figure}[!htb]
  \centering
  \includegraphics[width=\linewidth]{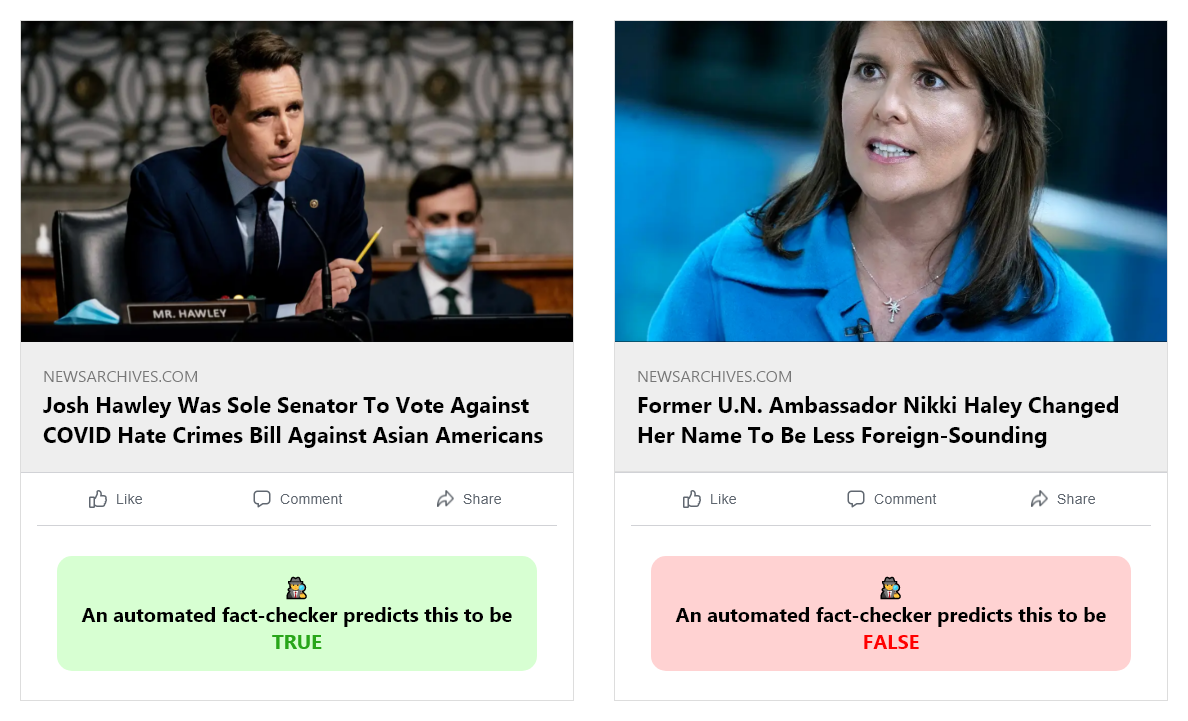}
  \caption{News that received a TRUE or FALSE prediction by the automated fact-checker. These visualizations were used as Control for our experiment.}
  \label{fig:NewsPrediction}
\end{figure}

For the study, 12 news on political topics were used with an equal ratio of true to false news.
The news were sourced from Snopes.com, a fact-checking service that uses an extensive rating system to cover the various dimensions of misinformation, within three months of the study. We only used news that Snopes rated to be either `True' or `False'~\cite{Snopes}. We then searched for the original news headlines from a variety of news websites and made grammar and case modifications to the headlines for a consistent reporting style. To prevent the source of news from influencing credibility perceptions of the content~\cite{Winter2014}, we chose a plausible yet neutral-sounding source domain for all the news: `newsarchives.com', that at the time of the study was not an active site. An interface similar to the Facebook news feed post that includes a label for the veracity prediction of the automated fact-checker was used for the news (Figure~\ref{fig:NewsPrediction}). This also served as the control.

\subsubsection{XAI Interfaces}
\label{sec:xaicond}
Participants were shown the XAI interface based on their assigned condition. The five XAI conditions were: LIME, SHAP, Attention, Comments, and Evidence. Each XAI interface contained a news (Figure~\ref{fig:NewsPrediction}) and a fact-checker report with a "How to understand this?" guide describing various aspects of the explanations (Figure~\ref{fig:XAI}). Larger individual images of the XAI interfaces can be found in Appendix~\ref{sec:XAIlarge}.

With the variety of conditions, we carefully considered the design of the fact-checker reports to ensure homogeneity in them and created mockups of the fact-checker reports instead of building actual XAI models to do so. Due to the marked differences in the visualization and mechanism of the conditions, we designed for homogeneity across graphical explanations: LIME (see Figure 14 in \cite{Ali2021}), SHAP (see Figure 7 in \cite{Reis2019}) and Attention (see Table 5 in \cite{Hansen2019}), and across textual explanations: Comments (see Figure 1 in \cite{Tian2020}) and Evidence (see Figure 2 in \cite{Popat2018}). We aimed to follow the original design of each condition (as referenced) as closely as possible while doing so.

\paragraph{Selection of conditions}
To identify relevant papers, keyword searches were conducted in the ACM Digital Library and IEEE Xplore, the two largest computing literature databases. A three-part query using combinations of \{`false information', `misinformation', `disinformation', `fake news'\} AND \{`explain*', `interpret*'\} AND \{`artificial intelligence', `AI', `machine learning', `ML'\} returned 308 ACM and 1061 IEEE unique results on 10 July, 2021. Screening of the results were conducted with the criteria that the articles were (1) full papers, (2) about an XAI or IML technique, (3) for misinformation detection, and (4) using texts. We found 31 papers that fit the criteria. A thorough reading of them identified the attention mechanism~\cite{Borges2019, Redi2019, Aloshban2020, Cruz2020}, evidence from other news sources~\cite{Popat2017, Popat2018, Nguyen2018, Wang2020}, and SHAP~\cite{Reis2019, Framewala2020, Vargas2020} as more popularly studied forms of explanations. Other techniques like knowledge graphs, decision trees and syntactic patterns had only one to two relevant papers each. Additionally, LIME~\cite{Jeronimo2019, Ali2021} was selected due to its generic popularity in XAI while explanations using social media comments~\cite{Cui2019, Tian2020} was included due to its similarity to evidence from other news sources. This led to the selection of the five XAI conditions.

\paragraph{Graphical XAI Interfaces}
For these interfaces, the automated fact-checker makes its prediction by weighing each unit of word in a line of text and providing a graphical representation of the contribution of each word as an explanation. For homogeneity, we designed all three XAI to consider only the words in the news. In doing so, we can make it such that the scores and the effects of the words were the same. For example, if the `fake' prediction probability in LIME is 0.26, the percentage of red arrows in the arrow bar in SHAP would be 26\%, and Attention would have a prediction score of 0.26. The effect of the words is treated in a similar fashion where words with greater effect have deeper shades or longer bars. As such, these conditions are also content-based as they explain using the news itself.

\paragraph{Textual XAI Interfaces}
For these interfaces, the automated fact-checker makes its prediction by weighing information from secondary sources, i.e., social media feeds for Comments and other news articles for Evidence, and uses the top matches as explanations. Since we used real comments from Twitter and Reddit and real articles from a variety of news websites to more truthfully emulate the XAI, it was largely impossible to get comments and evidence that paralleled each other. Therefore, for homogeneity in design, we focused on the visual aspect of the explanations by using the same layout (see Figure 1 in \cite{Tian2020}): a description at the top followed by three boxes that each have a score at the bottom right corner, using similar color schemes to indicate the stance of the boxes, and always having a box that either supports or is neutral towards the news as the top-most one. As such, these conditions are also context-based, as they explain via alternative information surrounding the news.

\subsection{Variables}

\paragraph{Independent Variables}
The first independent variable is the \cond. This independent variable is between-subjects: participants are exposed to only the LIME, SHAP, Attention, Comments or Evidence condition (see Figure~\ref{fig:XAI}). We also included a Control condition that only includes an indication of whether the headline is true or false (see Figure~\ref{fig:NewsPrediction}). All five XAI conditions are described in section~\ref{sec:xaicond}.

The second independent variable is the veracity of the news (\vera) and is within-subjects with each being either `True' or `False'. We assess this variable as an interaction with \cond.

\paragraph{Dependent Variables}

The first dependent variable is the agreement with the fact-checker (\agr). Participants were asked “How accurate do you think the news above is?” on a 4-point Likert scale \{1: False, 2: Somewhat False, 3: Somewhat True, 4: True\}. This did not have a neutral statement as we wanted the participants to make a decision on the veracity of the news, following prior literature assessing misinformation interventions~\cite{Clayton2020, Kirchner2020, Pennycook2019}. We then converted the perceived veracity into an agreement score by comparing it to the actual veracity of the news. A participant answering `True' on a real news (resp. `False' on a fake news) would reach an agreement score of 4, `Somewhat True' (resp. `Somewhat False') a score of 3, and so forth. We also compared the variances of the agreement score as a proxy to measure the reliance of users on the XAI when forming a veracity judgment.

The second dependent variable is the intent to share the news (\sha) through “I would consider sharing this news on social media.". This variable has been examined in prior literature due to its direct influence on the spread of misinformation~\cite{Clayton2020, Kirchner2020, Pennycook2019} and is on a 5-point Likert scale \{1: Definitely Not, 2: Somewhat Not, 3: Neutral, 4: Somewhat Yes, 5: Definitely Yes\}.

\subsection{Iterations of the Study Design}

Through several pilot studies, we refined the study by rephrasing the questions for better clarity, adding a progress indicator, and reducing the number of news shown. Realizing that participation fatigue was a major concern in our design due to the repetitiveness of the task and the cognitive burden in processing the fact-checker reports, we reduced from an initial set of 24 to 12 news. This reduction allowed for a more manageable completion time of around 15 minutes. While this led to less observations, we considered the tradeoffs such as abandonment of the experiment or lower quality responses in later trials due to boredom or distraction~\cite{Ganassali2008} and agreed on the reduction.

\subsection{Participants and Recruitment}
In reviewing the collected data, the exclusion criteria included (1) failing the screener survey, (2) failing the attention check, (3) completing the study within an exceptionally short or long time, (4) responding randomly during the study, and (5) searching online for more information during the study, Criteria 4 and 5 were assessed through respective Yes/No questions set at the end of the study. Of the 222 responses received, 42 were excluded for the above reasons.

The study had 180 participants. Their age ranged from 23 to 73 years old ($M=38.4, SD=10.5$) and a majority identified as male (57.8\%). On political ideology, a majority identified as democrats (62.8\%). On education, a majority had bachelor's degrees (55.0\%).
Through 5-point Likert scale questions (1: Definitely Not, 5: Definitely Yes) on news reading habits, participants reported reading news on social media (M=3.96, SD=1.10) and traditional news media (M=3.81, SD=1.29), and following fact-checked content (M=3.41, SD=1.15).

We adopted a rolling recruitment process until 30 accepted participants per condition were reached. On MTurk, workers had to be located in the US, with a HIT approval rate of at least 98\%, and more than 5000 approved HITs. Custom qualifications were used to prevent repeated attempts. Through the screener survey, participants were assessed to (1) be 18 years old and above, (2) live in the US for the last 5 years, (3) be fluent in English, (4) actively use and (5) share content on social media, and (6) agree to participate.

We gathered a total of 30 (participants) $\times$ 6 (XAI conditions) $\times$ 12 (news) = 2160 trials.

\subsection{Data Analysis}

\paragraph{Statistical Analyses for the Experiment}
Since the data was not normally distributed, a non-parametric approach was used for the data analysis.
For \cond, Kruskal-Wallis tests were used for main effects followed by pairwise Mann-Whitney U tests for pairwise comparisons.
We used Fligner-Killeen's test to compare the variance of the \agr\ scores across the XAI conditions and used the same for the pairwise comparisons.
Finally, aligned rank tests~\cite{Wobbrock2011} were used for interaction effects between \cond\ and \vera\ followed by contrast tests~\cite{Elkin2021} for pairwise comparisons.
Benjamini-Hochberg corrections were applied in all pairwise comparisons to minimize false discovery rates~\cite{Glickman2014}.

\paragraph{Qualitative Coding for the Post-Experiment Survey}
We conducted focused coding~\cite{Esterberg2002} on participants' responses to their perceptions of the assigned XAI interface. Through initial rounds of open coding, a codebook (Table~\ref{tab:afc}) was developed which was subsequently applied to the responses to capture common threads in them. Training of the coders occurred over three iterations, where disagreements were resolved through discussion and the definitions and protocols of the codebooks were refined. Two coders were trained on a random sample of 20\% of the responses, reaching an average interrater reliability of $\kappa=.89$ ($SD=.11, Min=.75, Max=1$), using Cohen's Kappa~\cite{Cohen1960}.

\section{Results: Experiment}

We describe the results of the experiment, addressing RQ1: What are the effects of different explanations on users' agreement with the fact-checker on the veracity of news? and RQ2: What are the effects of different explanations on users' intent to share news?

\subsection{Addressing RQ1: Effects of Explanations on User's Agreement with the Fact-Checker}

\begin{figure}[!htb]
    \centering
    \begin{subfigure}[b]{\linewidth}
        \centering
        \includegraphics[width=\linewidth]{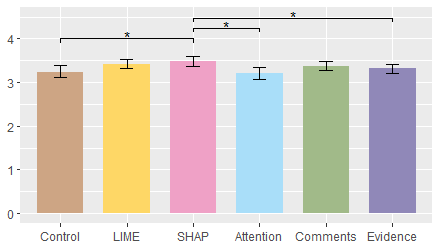}
        \caption{Agreement by XAI Condition}
        \label{fig:XAIagree}
    \end{subfigure}
    \hspace{0.05\linewidth}
    \begin{subfigure}[b]{\linewidth}
        \centering
        \includegraphics[width=\linewidth]{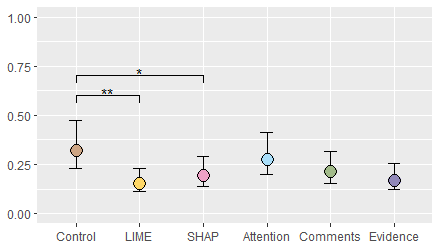}
        \caption{Variance by XAI Condition}
        \label{fig:XAIvar}
    \end{subfigure}
    \caption{Chart (a) shows the effects of XAI Condition on Agreement. Agreement scores are between 1 (strong disagreement) and 4 (strong agreement) with the averages shown. Chart (b) shows the variance in the agreement scores for each condition. Error bars show .95 confidence intervals. *: $p$<.05, **: $p$<.01.}
\end{figure}

There was a main effect of \cond\ on \agr\ ($\chi^2(5)=14.58, p=.012, \eta^2=.027$). Participants in the SHAP condition ($M=3.49, SD=0.44$) had significantly higher agreement scores than the Control ($M=3.24, SD=0.57$) and the Attention ($M=3.21, SD=0.53$) conditions (both $p<.05$). We also found a significant difference between SHAP and Evidence ($M=3.31, SD=0.41, p<.05$) but as discussed in section~\ref{sec:faircomparison}, comparisons across XAI categories can be problematic. Results are shown in Figure~\ref{fig:XAIagree}. 

The variances of \agr\ across \cond\ were not homogeneous ($\chi^{2}(3)=12.13, p=.033$) as shown in Figure~\ref{fig:XAIvar}. Both LIME ($SD=0.39, IQR=.33$) and SHAP ($SD=0.44, IQR=.50$) had significantly lower variances (both $p<.05$) than the Control ($SD=0.57, IQR=.67$). No difference between the \emph{Textual} XAI interfaces and the Control were found.

\subsubsection{Interaction Effects on Agreement}

We found a significant \cond\ $\times$ \vera\ interaction on \agr\ ($F_{5,174}=4.13, p=.0014$, $\eta^{2} = 0.12$).
Participants in the Attention condition had significantly lower agreement on true news than false news ($p<0.05$).
There was also an interesting observation with Evidence where the agreement was higher for true news than false news compared to all the other conditions where the agreement for false news was higher than true news.

 \begin{figure}[!htb]
    \centering
    \includegraphics[width=\linewidth]{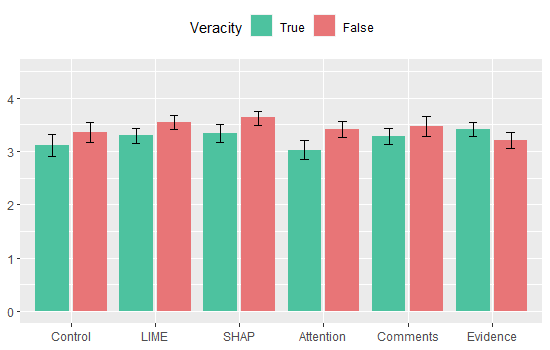}
    \caption{Interaction of XAI Condition and News Veracity on Agreement. Scores are between 1 (strong disagreement) and 4 (strong agreement) with the averages shown. Error bars show .95 confidence intervals.}
    \label{fig:XAIinteract}
\end{figure}

\subsection{Addressing RQ2: Effects of Explanations on User's Intent to Share News}

No main effect of \cond\ on the \sha\ was found ($\chi^2(5)=5.50, p=.36, \eta^2=.0014$). The descriptive statistics for each XAI condition is shown in Table~\ref{tab:XAIshare}. While the LIME, SHAP and Comments conditions seemed to lower the intent to share, this was not observed for the Attention and Evidence conditions. We also did not find an interaction effect with \vera\ ($F_{5,174}=1.34, p=.25, \eta^{2} = 0.037$).

\begin{table}[!htb]
  \caption{Descriptive statistics ($M (SD)$) of the XAI conditions on the intent to share. Scores are between 1 (definitely not) and 5 (definitely yes).}
  \label{tab:XAIshare}
  \begin{tabular}{ccccccc}
    \toprule
    & Control & LIME & SHAP\\
    \midrule
    Intent to Share & 2.28 (1.00) & 2.08 (1.02) & 2.04 (1.04)\\
    \midrule
    & Attention & Comments & Evidence\\
    \midrule
    Intent to Share & 2.29 (1.00) & 2.13 (0.87) & 2.29 (0.92)\\    
  \bottomrule
\end{tabular}
\end{table}

\section{Results: Post-Experiment Survey}
We describe the results of the post-experiment survey, addressing RQ3: What are the preferences towards explanations in automated fact-checking?

\subsection{Perceptions towards the Assigned XAI Interfaces}

For the open-ended question on the opinions on the XAI condition assigned, we excluded the Control condition, thereby analyzing 150 responses by participants in the five XAI conditions. Of these, 17 responses that were empty, not meaningful or irrelevant were discarded. In total, 133 responses with an average of 122.7 characters ($SD=84.2, Min=6, Max=396$) were analyzed. A summary of the codebook and results are shown in Table~\ref{tab:afc}. We coded the general sentiment and five qualities of the XAI fact-checker: usefulness, ease of understanding, accuracy, trustworthiness and design. The categories are further split into subcodes: positive, negative, conditional and mixed. Conditional refers to aspects that will be fulfilled only when some condition is satisfied and is thus neither positive nor negative. Mixed refers to aspects containing combinations of positive, negative and conditional elements.

The qualities are described as follows:
\begin{itemize}
    \item Usefulness - Level of aid provided by the fact-checker to help users ascertain the veracity of news
    \item Ease of Understanding - Effort required by users to interpret and comprehend the explanations
    \item Accuracy - Performance of the fact-checker in consistently making a correct decision on the veracity of news
    \item Trustworthiness - Reliability of the features used to build the algorithm and the ownership of the fact-checker
    \item Design - Visual appeal and organization of the explanations
\end{itemize}

\begin{table*}
  \caption{Codebook on the opinions on the assigned XAI condition ($N=133$). There are 6 code categories and 19 subcodes with the frequencies and interrater reliability values reported.}
  \label{tab:afc}
  \begin{tabular}{ccccc}
    \toprule
    Code & Subcode & Frequency ($n$) & Frequency ($\%$) & IRR ($\kappa$) \\
    \midrule
    \midrule
    \multirow{4}{*}{Sentiment} & Positive & 69 & 51.9 & \multirow{4}{*}{.75} \\
                      & Negative & 36 & 27.1 & \\
                      & Conditional & 7 & 5.3 & \\
                      & Mixed & 21 & 15.8 & \\
    \midrule
    \multicolumn{5}{c}{Qualities of the XAI Fact-Checker} \\
    \midrule
    \multirow{3}{*}{Usefulness} & Positive & 46 & 34.6 & \multirow{3}{*}{.93} \\
                      & Negative & 9 & 6.8 & \\
                      & Conditional & 4 & 3.0 & \\
    \midrule
    \multirow{3}{*}{Ease of Understanding} & Positive & 23 & 17.3 & \multirow{3}{*}{.88} \\
                      & Negative & 14 & 10.5 & \\
                      & Conditional & 0 & 0.0 & \\
    \midrule
    \multirow{3}{*}{Accuracy} & Positive & 26 & 19.5 & \multirow{3}{*}{1.0} \\
                      & Negative & 9 & 6.8 & \\
                      & Conditional & 2 & 1.5 & \\
    \midrule
    \multirow{3}{*}{Trustworthiness} & Positive & 7 & 5.3 & \multirow{3}{*}{.75} \\
                      & Negative & 23 & 17,3 & \\
                      & Conditional & 4 & 3.0 & \\
    \midrule
    \multirow{3}{*}{Design} & Positive & 19 & 14.3 & \multirow{3}{*}{1.0} \\
                      & Negative & 8 & 6.0 & \\
                      & Conditional & 0 & 0.0 & \\
    \bottomrule
  \end{tabular}
\end{table*}

\subsection{General Positive Opinions}
Participants had a largely positive opinion for the different XAI conditions as reported in Table~\ref{tab:afc}. Forty-six participants found the automated fact-checker useful or helpful with some participants noting that beyond fact-checking, the explanations made them consider the news more critically and be more cautious. P144 mentioned that the fact-checker \textit{"helps me look at news stories with an open-mind, and with less bias. I'm less likely to assume something is true, only because I want it to be true"}.
P164 further commented on autonomy saying that \textit{"I like that it gives people a way to determine if something is true or false, letting the person make that decision based on the facts given."} For the 23 participants who found the fact-checker easy to understand, they described it as \textit{"intuitive"} (P105), \textit{"straight forward"} (P175), \textit{"clear and concise and very easy to read"} (P70). Twenty-six participants found the fact-checker accurate and seven found it trustworthy with P139 saying that it \textit{"seems very reliable and I agreed with most of what the [fact-]checker had determined to be real versus fake."} These aspects could have been evaluated through the participants' personal understanding of the news as mentioned by P147 who said that \textit{"I recognized a few of the news stories and felt [that] the fact-checker was correct in it's assessment of them."} For the 19 participants who positively commented on the design of the fact-checker, they described it as \textit{"attractive"} (P44), \textit{"simple and clean"} (P38), with a \textit{"well structured format"} (P98) and \textit{"colors [that] were cool to look at"} (P149). Additionally, several participants suggested that they would want to have the XAI fact-checker in their everyday life. As put by P176, \textit{"I think that tools like this one will become increasingly necessary in the age of social media and rapid dissemination of misinformation."}

\subsection{General Negative Opinions}
Most of the negative opinions centered on the dependability of the fact-checker with 23 participants finding it untrustworthy.
First, there were doubts on AI technology itself. P154 said \textit{"I am always a bit `leery[]' with any information that is generated by an algorithm, because the algorithms are made by humans, and humans are fallible"}.
Second, the suitability of automating the fact-checking process was questioned with P49 saying that \textit{"I do not think [AI] fact checkers should ever[] be use[d] as they simply cannot understand context or be programmed in a way that can reliably fact check sources."} Lastly, a common concern with several participants was the manipulation of fact-checkers (both automated and not) to serve political purposes. Fact-checkers were thought to be \textit{"heavily biased propaganda machines meant to push a narrative"} (P48).
Strong defiance was expressed in these cases, with P108 asking \textit{"Who is fact checking the fact checkers[?] I can make up my own mind and don[']t need communist [sic] telling me what to think."}

As for the other qualities, nine participants felt that the fact-checker was not useful or needed. For P57, the fact-checker \textit{"created more doubts for me when it was in the middle of true or false!!"} while P1 said that \textit{"I investigate any news that is interesting to me on my own. I really do not need a fact checker."} Fourteen participants had difficulty understanding the fact-checker, finding the explanations confusing. P9 stated that \textit{"I would likely do additional research on articles before I share them at least 75\% of the time"} as a result. The nine participants who found the fact-checker inaccurate noted that there were errors made by the fact-checker such as there being \textit{"some things that were off with some of the stories that were definitely true"} (P153). As for the eight participants who commented negatively on the design, they felt that the explanation was \textit{"cumbersome"} (P92), \textit{"very cluttered [and] needs to be simplified"} (P96).

\subsection{Opinions by the XAI Condition}

\begin{figure}[!htb]
  \centering
  \includegraphics[width=\linewidth]{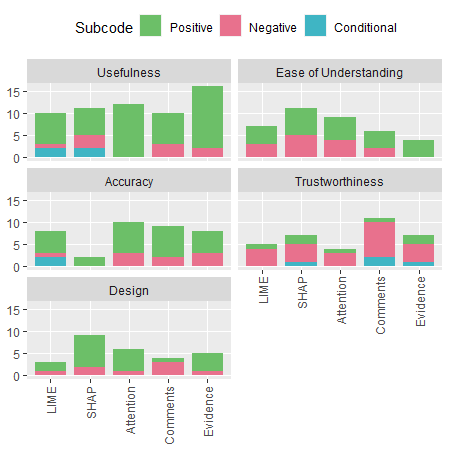}
  \caption{Frequencies (n) of the subcodes for the qualities of each XAI condition.}
  \label{fig:FreqAFC}
\end{figure}

Figure~\ref{fig:FreqAFC} shows the breakdown of the qualities mentioned for each assigned XAI condition. We further provide some quotes for each condition.

For LIME, there were contradicting opinions with some participants like P96 commenting that \textit{"it was very cluttered [and] needs to be simplified"} yet others like P118 saying that it had a \textit{"Simple and easy to use design."} The higher number of visuals could also have led P30 to comment that \textit{"I think that it's accurate, but I find the way it explains how it comes to its conclusions to be a bit confusing."}

There were similar observations for SHAP with P159 saying that \textit{"I did not fully understand how the fact checker mad[e] its determinations"} yet P29 commenting that \textit{"I thought it was a good tool and the graph was helpful in understanding its conclusions."} These were likely due to a learning curve in understanding the explanations. As put by P3, \textit{"I was a bit confused by how the fact checker worked, but was able to decipher it as the task went along"}.

For Attention, some positive comments included finding it \textit{"simple"} (P67) with a design that was \textit{"pretty good and easy to follow"} (P133) whereas some noted that it \textit{"did not highlight some important keywords in the news"} (P152) and that \textit{"the prediction score needs to be worked on as it can be confusing"} (P167).

For Comment, P67 said that \textit{"It's graphic design is so-so but the content is trustworthy and compelling."} although P119 instead remarked that \textit{"I am under the impression that the fact-checking comes from the comments? Which seem[s] completely biased and unreliable.. makes no sense!"}

For Evidence, P90 said that \textit{"The snippets it gave to read were helpful."} and P79 commented that \textit{"I like the layout and ease of use of it; it prompted me to think about the content I was consuming and made me question the articles."} However, there were also concerns about the source with P14 remarking that \textit{"This is a difficult question in a divided nation where propaganda, lies, and innuendo govern any "news" produced by the conservative right."}

In general, there were polarizing perceptions towards the XAI conditions in both their graphical and textual design and to the content and contextual information presented for the fact-checking.

\section{Discussion}

\subsection{Differences in Agreement Between the XAI Conditions}
\label{sec:shapvsattention}
The XAI conditions showed mixed results on the level of agreement with the fact-checker.

\paragraph{Comparing to the Control}
Of all the XAI conditions, only SHAP led to significantly higher levels of agreement.
This suggests that the presence of explanations itself may not strongly convince users of the automated fact-checkers's prediction.

\paragraph{Between Graphical Explanations: LIME, SHAP and Attention}
The Attention condition had lower levels of agreement to LIME and SHAP even though all three conditions were designed to give the same output.
The additional dimension of the polarity of the features in LIME and SHAP could have been more visually impactful and thereby more convincing compared to Attention which only had a single shade.
As such, apart from the algorithm used to perform the fact-checking, the way the results are explained to users is also important as design changes may lead to weaker persuasiveness of the XAI, and consequently, perhaps even lower levels of trust in them~\cite{Nourani2019}.

\paragraph{Between Textual Explanations: Comments and Evidence}
From the interaction effect with News Veracity, Evidence had higher agreement scores for true news and lower scores for false news, whereas Comments had higher scores for both, as compared to the Control. The observation of low agreement scores for false news in the Evidence condition is rather surprising. Were participants unconvinced by the excerpts from news articles? Or worse, were they instead convinced of the otherwise by them? Perhaps the weakening trust in media outlets is at play here~\cite{Reuters2022}. If so, this is a concerning observation that calls for a broader investigation. In contrast, while social media is not considered authoritative in the traditional sense, participants do seem to take to the content. This may be because social media comments tend to be more opinionated and written with expressive and emotional language that come across as more persuasive~\cite{VanKleef2015}.

\subsection{XAI Reduces Variance in Agreement}
Our analysis on the variance of the agreement scores showed that the variances of the five XAI conditions were lower than the Control despite their comparable mean scores, with the LIME and SHAP conditions being significantly so, possibly because they share similar features~\cite{Wang2021}.
A larger spread means a greater variation in the levels of agreement within a group of participants, and as such, the lower observed variance suggests that explanations, particularly LIME and SHAP, can help users determine whether a news is true or not more confidently. This is in line with the work of Lai and Tan~\cite{Lai2019} that found that "additional details including irrelevant ones can improve the trust that humans place on machine predictions" where trust was defined as "the percentage of instances for which humans follow the machine prediction". This demonstrates a clear benefit of explanations in that they are relied upon for determining the veracity of news and that they help in reinforcing veracity judgments, thereby leading to greater uniformity in the veracity perceptions.

\subsection{XAI Has Limited Impact on the Intent to Share}
As we did not observe any effect of the XAI condition on participants' intent to share, we conclude that the different conditions we chose for the study do not seem to affect participants' behavior in terms of sharing habits. This result suggests that automated fact-checkers may help users assess the veracity of news, but that incorporating XAI on a social network platform would not directly impact the number of shares of a given piece of news. Studies have shown that social factors are stronger motivators for sharing~\cite{Chen2013}. Thus, our observation is aligned with prior research.

\subsection{No Best Type of Explanation}
There were polarizing preferences towards the graphical/content-based and textual/context-based XAI interfaces among the participants from the post-experiment survey. This suggests that there is no perfect form of explanation for automated fact-checking, echoing the findings of works in other domains~\cite{David2021, Hamon2021, Wolf2020}. One's background, knowledge and experience shape one's being and this leads to the diversity of subjective preferences in people. This makes it challenging to settle on a single explanation as has been the case for user interface design in general~\cite{Carroll2003} which has led to the exploration of alternatives such as adaptive interfaces~\cite{Browne1990}. More investigations on examining and identifying explanations that are suitable or `good enough' for the contexts and goals of their use will be necessary.

\section{Implications for Design}

\subsection{Usability Considerations}
From the study, we surface some observations on the usability of XAI interfaces. First, content and visualization matters as incomprehensible or unconvincing explanations may lead to effects that are lower than or even opposite to what is expected. Second, with the widely varying preferences, providing a variety of explanations to enable users to have a comprehensive overview~\cite{Chromik2021} or tailoring an explanation to suit the needs of the user~\cite{Lim2022} would be more adequate than a single standardized explanation. Third, for the context-based mechanisms, participants displayed greater trust in evidence and greater distrust in comments, thereby mirroring the same impression onto the fact-checker. With explanations revealing the features of the AI, more tactfulness should be taken in the selection of the sources and factors used in the algorithm. Fourth, users tend to draw parallels between the automated fact-checking explanations and their own verification processes, thereby being more or less favorable towards the explanations. A field of AI is contestability where users are involved in "surfacing values, aligning system design and use with context, and building legitimacy"~\cite{Vaccaro2019}. The automated fact-checker can be improved by enabling greater human agency to challenge and shape the algorithm towards meaningful joint decision-making. Fifth, with information processing being cognitively burdensome, the amount of explanations provided by the fact-checker should be kept down. This was exemplified by how we had to reduce the number of news presented to participants as a result of cognitive taxation. Efforts should be taken to design concise explanations that minimize cognitive load~\cite{Abdul2020} which is not a trivial task as explanations also have to be informative. More research in this area would be beneficial towards the public adoption of XAI in any field. Lastly, and in line with the previous point, the integration of explanations in social media apps should not clutter the interface, particularly on mobile phones that have limited screen space. A plausible design suggested by P177 is \textit{"I'd rather something small but eye catching that says true or false, or mixed, and ... a link that explains rather than such a large explanation under the article."}

\subsection{Prospective User Evaluation Framework}
From the open-ended question, we identified five qualities that were consistently brought up in the feedback. These are the usefulness, ease of understanding, accuracy, trustworthiness and design of the fact-checker. While our study was not centered on the development of a framework, we realized that these qualities revealed aspects of the XAI that mattered most strongly to users. As such, we posit that they may factor as a framework in the user evaluation of XAI. Such a framework would be beneficial in assessing explanations in general, such as in other highly consequential domains of healthcare and law, where XAI is used to inform users, including professionals and the general public, of the decisions made by the algorithms. By having user-generated principles to assess explanations, this aligns with and supports the human-centered design of XAI.
Langer et al. have listed 29 stakeholders' desiderata on explainability~\cite{Langer2021}, some of which is common with those we describe, namely Usefulness, Accuracy and Trustworthiness, and there are others like Confidence and Satisfaction where parallels may be drawn with the remaining qualities. While the qualities we identify are not nearly as comprehensive as theirs, we believe that these qualities are intuitively more important to users and can serve as a concise user evaluation framework.

In the interest of continuity, we offer a brief description of the qualities:
\begin{itemize}
    \item Usefulness - Level of aid provided by the XAI to help users meet a practical goal
    \item Ease of Understanding - Effort required by users to interpret and comprehend the explanations
    \item Accuracy - Performance of the XAI in consistently making a correct decision
    \item Trustworthiness - Reliability of the features used to build the algorithm and the ownership of the XAI
    \item Design - Visual appeal and organization of the explanations
\end{itemize}

This prospective framework is, nevertheless, a coarse one that needs further investigation.

\subsection{Reflecting on Explainable Automated Fact-Checking}
During the qualitative coding process, we encountered two statements that made us contemplate on the practicality of using explanations in automated fact-checking. The first was by P52 who said that \textit{"text analysis can be gamed"} and the second by P12 saying that \textit{"I'd be worried of its veracity [in the] long-term, especially if people find out how it works and can then purposefully word their posts to game the system."} Both participants were speaking of content-based explanations and they were concerned that over time, users could easily figure out how the fact-checker weighted words, thereby crafting their posts in favor of a positive veracity and devaluing the automated fact-checker. We felt that this was a critical observation and seek to have an expanded discussion on it.

With explanations exposing the decision-making process of the automated fact-checker, it also poses a risk for malicious actors to unravel and attempt to outwit the system. Such concerns have been echoed before with Twitter saying that \textit{"to ensure people cannot circumvent these safeguards, we're unable to share the details of these internal signals in our public API"} for their bot detection algorithm~\cite{Twitter2017}. This also raises the question of whether explanations are ready to be released to the public even if they are well developed, especially in the fields of surveillance, security and safety. And if so, what types of data and XAI techniques are suitable for public-facing systems that hold such risks? For explainable automated fact-checking, perhaps example-based or generated natural language explanations~\cite{Atanasova2020} are best whereas feature-based explanations that reveal the inner workings of the algorithm should be kept for internal use such as by the system architects and the fact-checking experts engaged to review detected misinformation.

\section{Limitations}
This study was run on MTurk with only US-based participants. As such, results may not generalize broadly to different countries and contexts.
Design-wise, the study was done with the XAI condition as a between-subjects factor in consideration of the study duration. Using a within-subjects design for this variable could have led to better comparisons but would likely have induced greater participation fatigue.
Also, as we examined only five types of XAI conditions widely used in the literature, further studies may want to investigate new designs or existing ones that were not included.

The study was done while explainable automated fact-checkers are not in the mainstream which might raise doubts on its applicability. Yet, there are already indications of eager advances to use automated fact-checking to aid in the work of human fact-checkers and to potentially expose them to users with the turn towards Responsible AI by social media firms. The perceptions of AI and fact-checkers may continue to evolve in the future and this work paves a path towards that understanding by investigating explanations that are at the fore of current explainable misinformation detection research. Furthermore, beyond the context of fact-checking, several of the considerations surfaced in this study are on the explanations themselves, rather than the context they are applied in, which are generic and applicable in other domains.

\section{Conclusion}
To understand the behavioral effects of explanations and how they are perceived by users, we conducted a study on five XAI interfaces in the field of misinformation where efforts have been made to improve the explainability of automated fact-checkers. We found various effects on users' veracity perception of news and their intent to share them by the XAI condition and the veracity of the news. In particular, we observed that XAI has mixed or limited effects on the agreement with the fact-checker and the intent to share news, two aspects that are important performance indicators of the effectiveness of fact-checking. Despite so, there is a reliance on explanations in the veracity judgments of news as shown by the decrease in agreement variances when compared to the control. As such, we conclude with the first half of our title: while not in the key areas, XAI does offer modest benefits in automated fact-checking. Our participants also displayed polarizing preferences between the graphical/content-based and textual/context-based explanations, with some raising concerns on the effectiveness, reliability and usability of the XAI fact-checkers. Thereby, we conclude with the second half of our title: that there is no single type of explanation that suits all. In light of this, we consolidated several considerations and suggestions on making explanations more practicable in both the context of fact-checking and in general. We also put forward a prospective user evaluation framework for explanations that was formulated from participants' feedback on the XAI interfaces. With explainability coming back to the fore in recent years, we aim for this work to contribute to more human-centered practices in the design of XAI. As future work, we plan to conduct more studies to understand the effects we have observed and to characterize the user types for the varying explanation preferences.

\bibliographystyle{ACM-Reference-Format}
\bibliography{main}

\newpage

\appendix

\section{XAI Interfaces}
\label{sec:XAIlarge}
\begin{figure}[H]
  \centering
  \includegraphics[width=.8\linewidth]{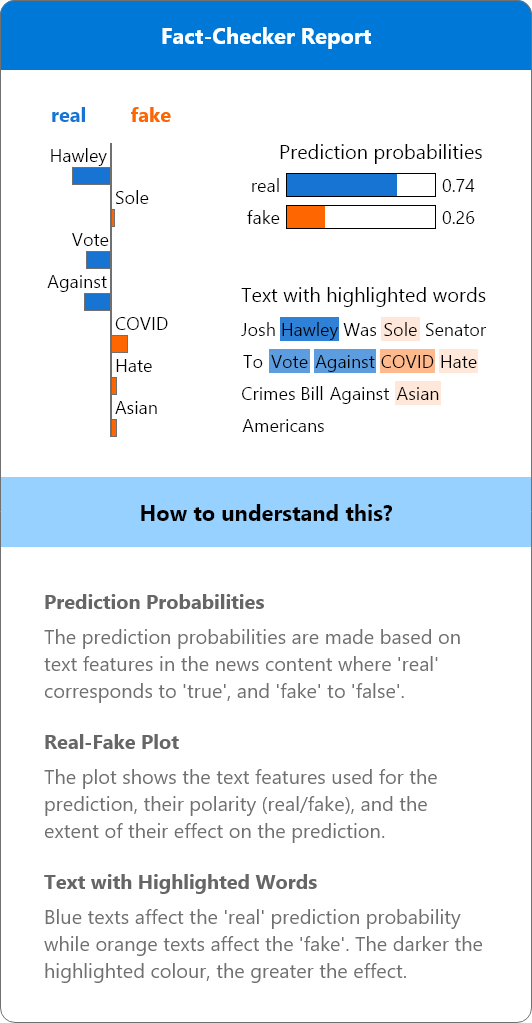}
  \caption{The LIME explanation.}
  \label{fig:XAI_LIME}
\end{figure}

\begin{figure}[H]
  \centering
  \includegraphics[width=.8\linewidth]{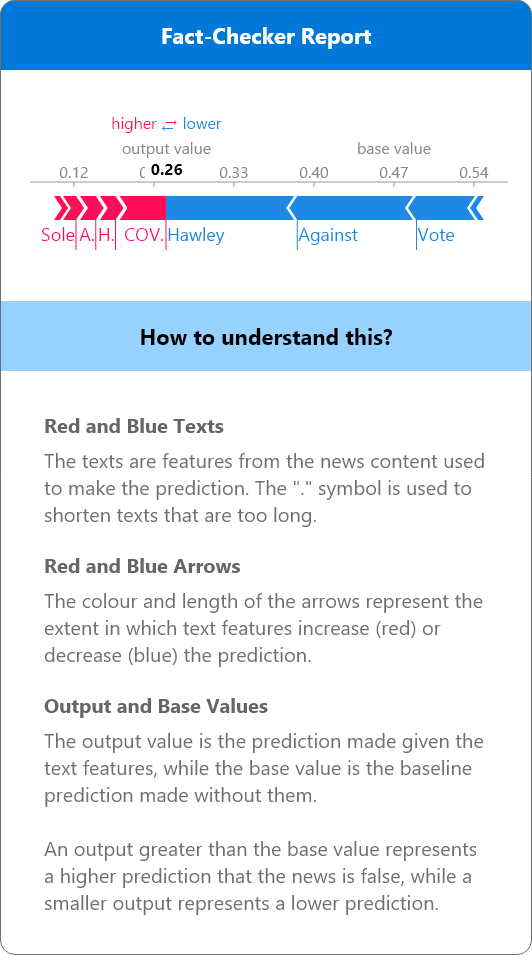}
  \caption{The SHAP explanation.}
  \label{fig:XAI_SHAP}
\end{figure}

\begin{figure}[H]
  \centering
  \includegraphics[width=.8\linewidth]{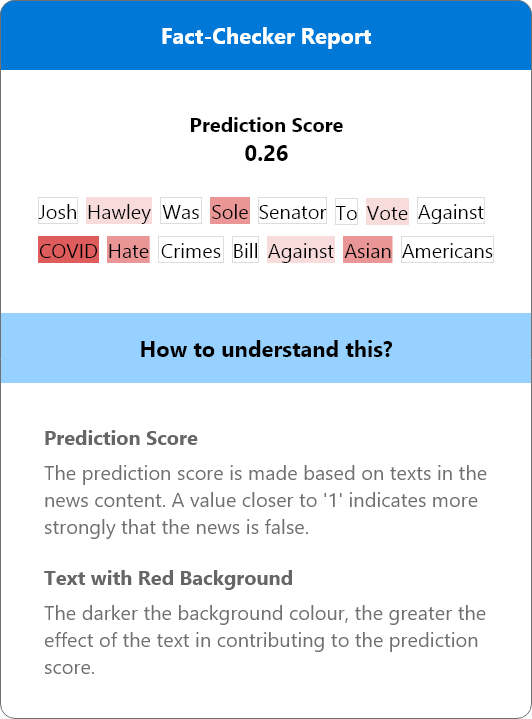}
  \caption{The Attention explanation.}
  \label{fig:XAI_Att}
\end{figure}

\begin{figure}[H]
  \centering
  \includegraphics[width=.8\linewidth]{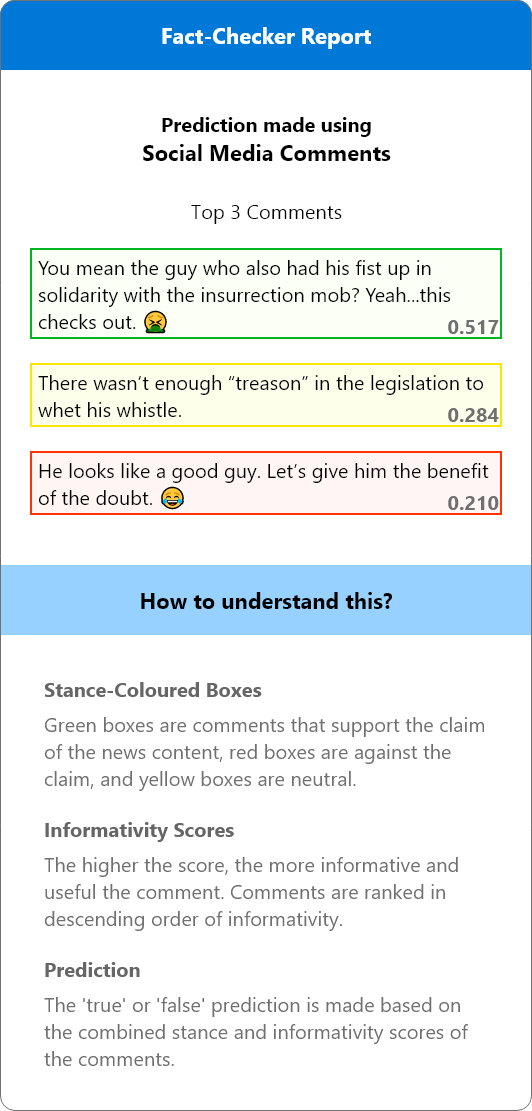}
  \caption{The Comments explanation.}
  \label{fig:XAI_Com}
\end{figure}

\begin{figure}[H]
  \centering
  \includegraphics[width=.8\linewidth]{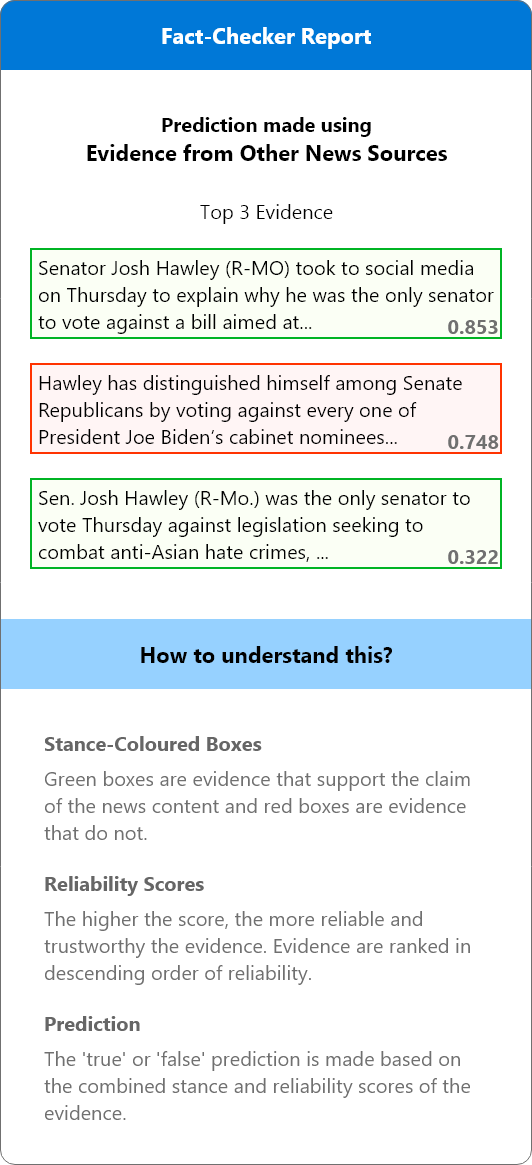}
  \caption{The Evidence explanation.}
  \label{fig:XAI_Evi}
\end{figure}

The five XAI interfaces: LIME (Figure~\ref{fig:XAI_LIME}), SHAP (Figure~\ref{fig:XAI_SHAP}), Attention (Figure~\ref{fig:XAI_Att}), Comments (Figure~\ref{fig:XAI_Com}), and Evidence (Figure~\ref{fig:XAI_Evi}).

\newpage

\end{document}